\documentclass[modern]{aastex62}
\usepackage{soul}
\usepackage{xcolor}

\usepackage{placeins}

\usepackage{float}

\shorttitle{C/2016 R2: A comet rich in CO and depleted in HCN}
\shortauthors{Wierzchos and Womack}

\begin{document}

\title{C/2016 R2 (Pan-STARRS): A comet rich in CO and depleted in HCN}

\correspondingauthor{K. Wierzchos}
\email{kacperwierzchos@gmail.com}

\author[0000-0002-0786-7307]{K. Wierzchos}
\affiliation{University of South Florida \\
Department of Physics \\
4202 E. Fowler Ave \\
Tampa, FL, 33620 USA}

\author[0000-0003-4659-8653]{M. Womack}
\affiliation{University of South Florida \\
Department of Physics \\
4202 E. Fowler Ave \\
Tampa, FL, 33620 USA}



\begin{abstract}

We observed comet C/2016 R2 (Pan-STARRS) with the ARO 10-m SMT, and report the first detection of CO emission from this comet with amounts high enough to be the primary driver of activity. We obtained spectra and maps of the CO J=2-1 rotational line at 230 GHz between 2017 December and 2018 January. We calculated an average production rate of Q(CO)=(4.6$\pm$0.4)x10$^{28}$ mol s$^{-1}$ at $r \sim$ 2.9 au and $\Delta \sim$ 2.1 au. The CO line is thin ($\Delta V_{FWHM} \sim$ 0.8 km s$^{-1}$) with a slight blue-shift ($\delta v$ $\sim$ -0.1 km s$^{-1}$) from the ephemeris velocity, and we derive a gas expansion velocity of $v_{exp}$ = 0.50 $\pm$ 0.15 km s$^{-1}$. This comet produced approximately half the CO that comet C/1995 O1 (Hale-Bopp) did at 3 au. If CO production scales with nucleus surface area, then the radius need not exceed $R_{R2}$ $\sim$ 15 km. The spectra and mapping data are consistent with CO arising from a combination of a sunward-side active area and an isotropic source. For HCN, we calculated a 3-sigma upper limit production rate of Q(HCN) $<$ 8x10$^{24}$ molecules s$^{-1}$, which corresponds to an extraordinarily high abundance ratio limit of Q(CO)/Q(HCN) $>$ 5000. We inferred a production rate of molecular nitrogen of Q(N$_2$) $\sim$ 2.8x10$^{27}$ molecules s$^{-1}$ using our CO data and the reported N$_2$/CO column density ratio \citep{coc18,coc18b}. The comet does not show the typical nitrogen depletion seen in comets. The CO-rich, N$_2$-rich and HCN-depleted values are consistent with formation in a cold environment of $T < $ 50 K that may have provided significant N$_2$ shielding.
\end{abstract}




\bigskip
\bigskip
\bigskip
\bigskip
\section{Introduction} \label{sec:intro}
Comets, comprised largely of ice and dust, constitute the least processed bodies of the Solar System, and most travel around the Sun in eccentric orbits. Their nuclei contain well-preserved samples of grains and gas from the protosolar nebula cloud in which they formed \cite{mum11}. As a comet approaches the Sun, it forms a coma around the nucleus by sublimating volatile ices, which release dust and other gases. In order to constrain models of comet (and by extension solar system) formation it is vital to accurately determine the composition and physical state of cometary nuclei \cite{com97,coc99}. 

Comet C/2016 R2 (Pan-STARRS) -hereafter R2- was discovered at $r$ = 6.3 au from the Sun on 2016 September 7 when it exhibited a 20$\arcsec$ wide coma at 19.1 visible magnitude \citep{wer16}. It has an estimated orbital period of 20,000 years, a highly eccentric orbit tilted at an angle of 58 $\deg$ to the ecliptic, and a semi-major axis of $a \sim$ 740 au. These orbital characteristics identify the object as an Oort Cloud comet, but not dynamically new, since it has presumably already had many journeys through the inner solar system \citep{levi96}.

Upon discovery, this comet exhibited a coma at a distance where most comets appear inactive. Water is the dominant ice in all comets and is not heated enough by the Sun to sublimate efficiently until much closer, typically $r = 2 - 3$ au. Thus, comet R2 also receives the ``distantly active comet" classification. Instead of water-ice sublimation, the observed comae of distant comets are generally considered to be due to release of cosmogonically abundant hypervolatile species, such as CO and/or CO$_2$ \citep{oot12,rea13,bau15,wom17,wie17}.

By late 2017, optical images of R2 revealed a deep-blue-colored coma and ion-tail with an absence of dust. The blue color in the coma is largely due to emission from CO$^+$ (a photoionization production of CO) and to some extent, N$_2^+$, which were both observed to be strong, with a ratio of N$_2$/CO = 0.06, one of the highest ever reported for a comet \citep{coc18,coc18b}. The strong presence of CO$^+$, and the lack of CO$_2^+$ emission in these optical spectra indicate that the comet's activity is probably dominated by the outgassing properties of CO and not CO$_2$. The comet's high N$_2^+$ abundance is very important, because its likely parent, N$_2$ is typically depleted in comets, and like CO, N$_2$  sublimates at extremely low temperatures \citep{iro03,wom17}. The abundance of N$_2$ is high enough that it may play a substantial role in R2's distant outgassing behavior. Furthermore, N$_2$ is also an important molecule for astrochemical models of the solar system and other planetary systems \citep{feg89,lodd10,mos16}.

\section{Observations and Results} \label{sec:obs}

We used the Arizona Radio Observatory 10-m Submillimeter Telescope in order to search for CO J=2-1 (at 230.53799 GHz) and HCN J=3-2 (at 265.88643 GHz) emission in comet C/2016 R2 during 2017 December - 2018 January when its heliocentric and geocentric distances were $r \sim$ 2.9 au and $\Delta \sim$ 2.1 au, respectively (Table 1).

\begin{center}\textbf{Table 1:} Observations of Comet C/2016 R2 

\begin{tabular}{||c c c c c c||}
 \hline
 \label{tab:obs}
 Molecule & UT Date$^{1}$ & $r$ (au) & $\Delta$(au) & T$_A^*$dv (K km s$^{-1}$)  & Q (x10$^{28}$ mol s$^{-1}$)\\ [0.5ex] 
 \hline\hline
 CO(2-1) & 2017-12-22.21 & 2.98 & 2.05 & 0.26$\pm$0.01 &  4.4$\pm$0.2\\ 
 
   & 2017-12-23.09 & 2.97 &2.05&0.28$\pm$0.02  & 4.6$\pm$0.4 \\
   & 2017-12-30.14 & 2.94 & 2.06&0.26$\pm$0.02  & 4.6$\pm$0.4 \\
   & 2017-12-31.13 & 2.93 & 2.06&0.27$\pm$0.02  &4.6$\pm$0.4 \\
   & 2018-01-16.03 & 2.86 & 2.15&0.27$\pm$0.02  &4.6$\pm$0.4 \\
 \hline
 
 HCN(3-2) & 12-23.19 \& 01-16.15 & 2.92 & 2.10 & $<$0.030$^2$ & $<$ 0.0008\\
\hline

\end{tabular}
\label{tab:obs} 
\begin{flushleft}
\tablenotetext{1}{Dates listed represent the midpoint of data collection.}
\tablenotetext{2}{The T*$_A$dv line area upper limit for HCN was calculated using three times the rms of the HCN spectrum multiplied by an assumed linewidth of 1 km s$^{-1}$.  For HCN, $r$ and $\Delta$ are the average of the comet's distances on December 23, 2017 and January 16, 2018.}
\end{flushleft}

\end{center}

\bigskip

We used the dual polarization 1.3 mm receiver with ALMA Band 6 sideband-separating mixers for all observations. The mode of data acquisition was beam-switching mode with a reference position of +2$\arcmin$ in azimuth. An integration time of 3 minutes on the source and 3 minutes on the sky reference position for each scan was used. System temperatures were typically in the low 300 K for all the data. The temperature scale for all SMT receiver systems, T$_A^*$, was determined by the chopper wheel method, with T$_R$=T$_A^*$/{$\eta_b$}, where T$_R$ is the temperature corrected for beam efficiency and $\eta_b$  is the main beam efficiency of the SMT with a value of $\eta_b$ = 0.74 for both the CO and HCN frequencies. The backends consisted of a 2048 channel 1 MHz filterbank used in parallel (2 x 1024) mode and a 250 kHz/channel filterbank also in parallel (2 x 250). The 250 kHz/channel filterbanks provided the equivalent velocity resolutions of 0.325 km s$^{-1}$ for CO J=2-1 and 0.282 km s$^{-1}$ for HCN J=3-2. The 1 MHz resolution filterbanks were significantly broader than the expected CO and HCN linewidths and thus were not used in the analysis. The ARO SMT beam size is $\theta_B$ = 32.7$\arcsec$ at the CO J=2-1 frequency  and 28.4$\arcsec$ at the HCN J=3-2 frequency.

After every six scans on the comet, we updated the pointing and focus on Uranus and on the strong radio-source Orion-A. The pointing and tracking were showing an accuracy of $<$ 1$\arcsec$ RMS throughout the observing epochs. The comet's phase angle ranged from 7 degrees (on 2017 Dec 22) to 15 degrees (on 2018 Jan 16). CO emission was detected and remained relatively constant during this time (Table 1). 


 \begin{figure}[ht!]
\plotone{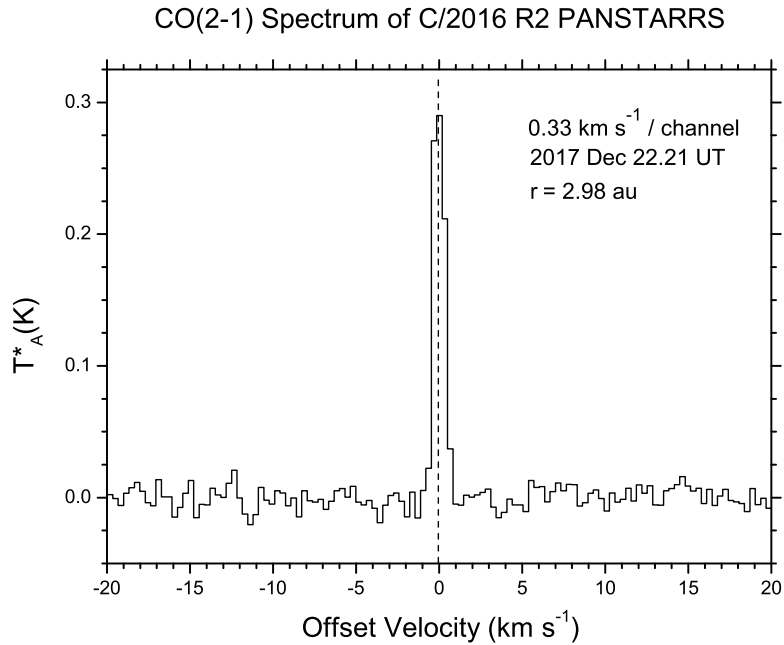}
\caption{The first detection of CO emission in C/2016 R2 on UT 2017 Dec 22, when the comet was at $r$ = 2.9 and $\Delta$ = 2.1 au. The CO J=2-1 rotational line was bright with an intensity of T$_A$* = 0.3 K easily detected in single scans with the Arizona Radio Observatory Submillimeter 10-m telescope. The spectrum was obtained with 250 kHz/channel spectral resolution with a Gaussian measured linewidth of FWHM = 0.79 km s$^{-1}$ and was not significantly shifted from the comet ephemeris velocity, which is indicated by a vertical dashed line at zero offset velocity.  \label{fig:original}}
%
%
%
\end{figure}

The CO J=2-1 line was detected in R2 during a single six minute scan on the  UT 2017 Dec 22 observations, and the total for the first day is shown in Figure \ref{fig:original}. The spectrum in Figure \ref{fig:original} is the first detection of CO emission in this comet, which we announced to the astronomical community in a preliminary report \citep{wie17b}. Both polarizations showed the line, and little change was observed in the line intensity, shape or area throughout the observing period. The $\sim$ 3 channel wide (FWHM) line profile yielded a $\Delta$V$_{FWHM}$ $\sim$ 0.79 $\pm$ 0.33 km s$^{-1}$ if a Gaussian fit is assumed. The line is typically blue-shifted by a small amount ranging from $\delta$v = -0.08 to -0.18 km s$^{-1}$.
 
\begin{figure}[H] 
\centering CO J=2-1 emission in comet C/2016 R2 
\plotone{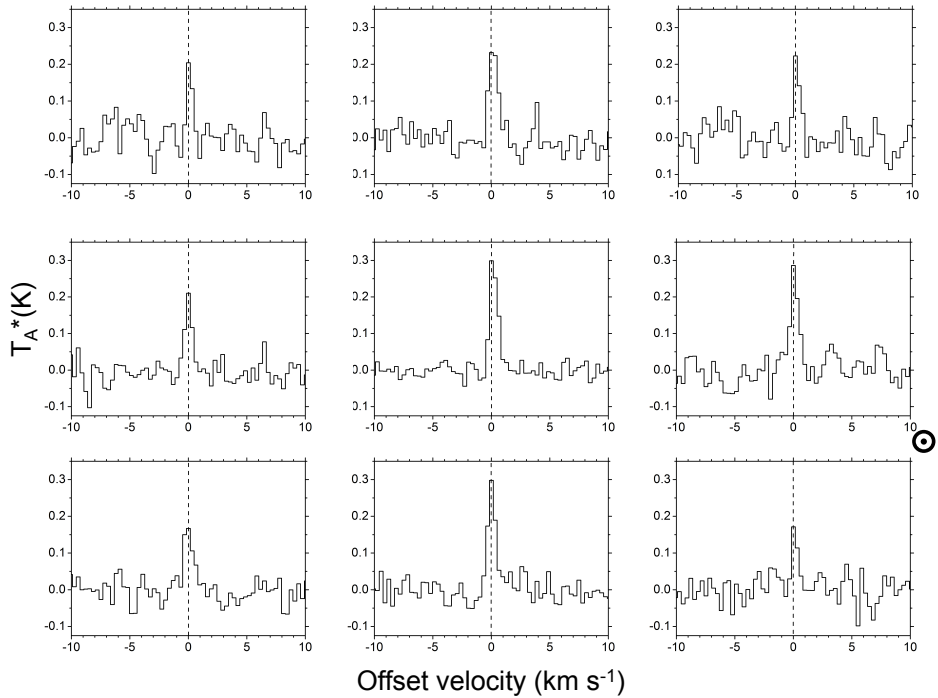}
\caption{Map of CO emission from C/2016 R2 constructed on 2018 January 16 with the ARO SMT. The size of the map corresponds to 96,000 km x 96,000 km (64\arcsec x 64\arcsec) on the sky at the comet's projected distance. The direction to the Sun is toward the right as indicated in the Figure. The comet's ephemeris speed is indicated with vertical dashed lines at zero velocity. CO emission peaked in intensity at the ephemeris location of the nucleus (center position) and may be slightly increased on the sunward side when compared to the tailward side. \label{fig:map}}
\end{figure}

We also mapped the CO emission on UT 2018 Jan 16.1 to assess its spatial extent in the inner coma. The map was constructed with a 9-point grid technique centered on the nucleus position. The map had 16\arcsec~spacings and integrations of 6 minutes for each position. The pointing separations for the map are equal to the half-power beamwidth (HPBW) of the SMT 10-m dish at this frequency (see Figure \ref{fig:map}). The map was aligned along the RA and Dec axis and the direction to the Sun is indicated in the figure. 

We searched for the HCN J=3-2 transition on 2017 Dec 23 and 2018 Jan 16 and did not detect a line down to a cumulative 1-sigma level of T$_A$* = 0.010 K in the 250 kHz/channel filterbanks (Table 1). We also searched for, and did not find, CH$_3$OH (251 GHz), H$_2$CO (218 GHz), N$_2$H$^+$ (279 GHz), HCO$^+$ (276 GHz) and CS (244 GHz). The significance of those non-detections and limits will be discussed in a later paper.

\section{Analysis and Discussion} \label{sec:discussion}

\subsection{CO outgassing velocities and spatial extent} \label{subsec:profile}

Important modeling parameters, such as expansion velocity, $v_{exp}$, and outgassing patterns, can be extracted from spectral line profiles and maps that have sufficiently high resolution \citep{biv02}. Comet activity models typically consider two different sources for CO in comets, one emanating from the subsolar point where solar heating is greatest, and one from an isotropic source in the coma. Here we briefly address how the R2 CO data aligns with the models.  

First, we examine the CO line profile, which has a single velocity component and is slightly blue-shifted from the comet's ephemeris velocity by $\delta v$ = -0.12 $\pm$ 0.20 km s$^{-1}$ (see Figure \ref{fig:original}). The line is between 2-3 channels wide (corresponding to FWHM of 0.66 - 0.99 km s$^{-1}$), and by fitting a Gaussian, we derived a FWHM linewidth of $\Delta v_{FWHM}$ = 0.85 $\pm$ 0.33 km s$^{-1}$. Thus, the data are consistent with having a FWHM linewidth of $\sim$ 0.8 km s$^{-1}$. The small velocity shift we observe of $\sim$ -0.1 km s$^{-1}$ for the comet with a low phase angle is consistent with at least some of the outgassing takes place on the Earth-facing side. The half-width half-maximum (HWHM) linewidth measured on the blue-ward wing is proportional to the outflow velocity of the gas (see \citet{biv99}), and thus, we estimate the gas expansion velocity to be $v_{exp}$ = 0.50 $\pm$ 0.15 km s$^{-1}$. This is comparable to what was measured in other comets at this same heliocentric distance, such as C/1995 O1 (Hale-Bopp) and C/2006 W3 (Christensen) \citep{gun03,boc10}. \cite{val18} also report seeing CO emission from this comet approximately three weeks after our first detection, but with a velocity redshift of $\sim$ +1 km s$^{-1}$ and a somewhat broader linewidth of 1.0-1.3 km s$^{-1}$. We also observed the comet during this time and do not confirm their redshifted velocity component.

The nearly-centered and narrow CO spectrum of R2 is consistent with a simple model of cool gas expanding isotropically with $v_{exp} \sim$ 0.50 km s$^{-1}$, which is what we used to calculate production rates. The data are also consistent with a more detailed scenario if one looks more carefully at the spectral line profile. First, we revisit Hale-Bopp and Christensen, which also produced substantial CO at $r$ = 3 -- 4 au. The CO emission in these comets were also slightly offset from the ephemeris velocity by $\delta v \sim$ -0.1 km s$^{-1}$ \citep{biv97,wom97,boc10}.  For Hale-Bopp, the CO line profiles were fit by a detailed two-component model comprised of isotropic outgassing of cold CO gas, combined with a blue-shifted velocity component associated with a sunward side active area \citep{gun03}. This model produces two peaked lines in comets beyond 4 or 5 au and single peaked lines for comets within 4 au. Therefore, the CO emission is also consistent with CO arising from a mix of a sunward-side active area and a symmetric source either in the coma or from the nucleus. 

We also examined the maps for clues about the CO outgassing. Figure \ref{fig:map} shows that the emission peaks at the nucleus position provided by the ephemeris. Furthermore, CO emission was readily detected at all positions out to at least 45\arcsec~ with a decrease in intensity by 20-40\% relative to the line at the center position, consistent with isotropic outflow of CO. There is evidence for a slight sunward enhancement of emission on the sunward side compared to the tailward side. This gives further support for contribution from an active area releasing CO on the sunward side, as is also indicated from the spectral line profile. 

It is not clear what could generate isotropic CO emission in the coma. Given that CO$_2^+$ was not detected in the optical spectrum of R2 \citep{coc18} and we did not see CH$_3$OH or H$_2$CO down to significant limits (Wierzchos, in prep.), it is not likely that CO was produced in significant amounts by photodissociation of these species, which are plausible secondary sources for CO in other comets. Also, this comet has an almost nonexistent dust coma, and thus CO is not likely to come from refractory cometary grains in the coma. Perhaps additional CO is released by sublimating water ice grains in the coma that were ejected from the nuclear sunward-side facing CO source. Measurements of OH or H$_2$O emission would be useful constraints to the CO production model in R2. Much higher resolution spectra, $\leq$ 0.1 km s$^{-1}$, would also be valuable for testing models of CO production.

In principle, more detailed modeling of the spectral line profile and mapping data could significantly constrain models of the release mechanisms for CO and physical conditions in the coma, but this requires higher spectral and spatial resolution data, and is beyond the scope of this paper.

\subsection{Production rates of CO and HCN} \label{subsec:production}

We calculated column densities assuming the excitation was dominated by collisional and fluorescence contributions, following the modeling described in \citet{cro83}, \cite{boc85}, and \citet{biv97}. We assumed a rotational and excitation temperature of 25K, which is consistent with the empirical fit to Hale-Bopp CO data described in \cite{biv02}. 

The column density for CO was fairly constant, with an average value of N(CO) = (1.89 $\pm$ 0.14)x10$^{14}$ cm$^{-2}$. In order to calculate production rates, we assumed a gas expansion velocity of 0.50 km s$^{-1}$, which is consistent with the CO spectral line profile and values from other comets at this distance, as described in Section \ref{subsec:profile}. Using a photodissociation decay model \citep{has57} and assuming isotropic outgassing of CO we find an average production rate of Q(CO) = (4.6 $\pm$ 0.4)x10$^{28}$ mol s$^{-1}$ between 2017 December 22 and 2018 January 16 (see Table 1). \cite{val18} report higher CO production rates, despite reporting similar line intensities. We think the different production rates are the result of using different modeling parameters, such as expansion velocity. There is insufficient detail about the modeling in their preliminary announcement to warrant further comments.

As discussed in section \ref{subsec:profile}, CO$_2$ is not likely to be a significant parent of CO in this comet. Infrared observations of CO$_2$ would be very useful in quantifying any contributions from CO$_2$ to the coma and/or CO emission. Also, to date, no searches for OH or H$_2$O emission have succeeded, which implies that water-ice sublimation is probably not responsible for most of R2's activity. Thus, the major driver at this distance is probably CO outgassing. 

The CO production rate of R2 is very high and approximately half that of C/1995 O1 Hale-Bopp at this same distance from the Sun. Based on the high Q(CO) values, we consider R2 to be ``CO-rich." Other CO-rich comets, typically have CO/H$_2$O $>$ 8\% \citep{dellorusso16} and the ratio for R2 may be substantially higher than 8\%, since water has yet to be detected. We point out that these observations were obtained when the comet was at $\sim$ 3 au from the Sun, which is too far for water-ice to sublimate efficiently. Thus, the relative CO/H$_2$O content in the nucleus may be much higher than we can determine at this heliocentric distance.

The nucleus' radius, $R_{R2}$, can be estimated based on the assumption that  Q(CO) is proportional to the nucleus surface area and the insolation received, and then compared to comets at the same heliocentric distance for which both Q(CO) and radius are independently known. For example, if we use R$_{HB}$ = 30 km \citep{fer02} and Q(CO) = 1.8x10$^{29}$ mol s$^{-1}$ for Hale-Bopp (at 3 au) \citep{biv02}, then the average Q(CO) = 4.6 x10$^{28}$ mol s$^{-1}$ at $\sim$ 3 au corresponds to a radius of $R \sim$ 15 km. Similarly, from a comparison with comet Christensen's radius upper limit of $R_{Ch} <$ 13 km \citep{kor16} and Q(CO) = 3.9x10$^{28}$ mol s$^{-1}$ \citep{boc10} at 3 au, we derived that $R  < $ 14 km. CO activity may not, in fact, scale directly with surface area for this comet, but if it does, then we find that the nucleus radius need not exceed $R_{R2}$ $\sim$  15 km in order to explain the measured CO production rate. 

We derived a 3-sigma upper limit of Q(HCN) $<$ 8.0x10$^{24}$ molecules s$^{-1}$ from all the HCN data. For comparison, this is $\sim$ 100 times lower than observed for Hale-Bopp at the same distance \citep{biv02}. Our non-detection of HCN emission is consistent with the absence of the CN band at 3880 Angstroms, as reported by \cite{coc18}, assuming that CN is caused by photolysis of HCN.

\begin{center}\textbf{Table 2:} Compiled Q(CO)/Q(HCN) ratios in CO-rich and other comets

 \begin{tabular}{||l r c r||} 
 \hline
 Comet & Q(CO)/Q(HCN) & $r$* (au) & Reference\\ [0.5ex] 
 \hline\hline
 \linethickness{0.05mm}
 C/2016 R2 (Pan-STARRS) & $>$5000 & 2.9 & This paper \\
 \hline
 
  29P/Schwassmann-Wachmann 1& 3300$^{\dagger}$ & 5.8 & [1,20] \\
  \hline 
  
 C/2006 W3 (Christensen) & 243 & 3.2 & [21] \\
 \hline
 
 C/1995 O1 (Hale-Bopp)  & 125-650 & 3 & [3] \\
 & 52-91 & 0.9 & [2,3,7]\\ 
 
\hline
 C/2010 G2 (Hill) & 70 & 2.5 & [16] \\

\hline

 C/1996 B2 (Hyakutake) & 96 & 0.6, 0.7 & [8] \\ 
 
\hline

 C/1999 T1 (McNaught-Hartley) & 46 & 1.3 & [9,11] \\
\hline
 C/2001 Q4 (NEAT) & 31 & 1.0 & [14] \\
 \hline

 C/2009 P1 (Garrad) & 36 & 1.6, 2.1 & [12,13,15,18] \\
 \hline

 C/2013 R1 (Lovejoy) & 34 & 1.3 & [17] \\

  \hline\hline
    Oort Cloud Comets & 28 &  - & [19] \\
     Jupiter Family Comets & 9 &  - & [19] \\
        All comets & 25 & - & [19] \\
        \hline

\end{tabular}
\label{tab:cohn}
\end{center}

References: [1] \cite{coch91}, [2] \cite{mag99},  [3] \cite{biv02}, [4] \cite{dis02}, [5] \cite{del02,del04}, [6] \cite{mag02}, [7] \cite{bro03}, [8] \cite{dis03}, [9] \cite{gib03}, [10] \cite{kaw03}, [11] \cite{mum03}, [12] \cite{pag12}, [13] \cite{vil12}, [14] \cite{val13}, [15] \cite{dis14}, [16] \cite{kaw14}, [17] \cite{pag14}, [18] \cite{mck15}, [19] \cite{dellorusso16}, [20] \cite{wom17}, [21] \cite{boc10} 

\medskip
\begin{flushleft}

*$r$ is the heliocentric distances at which production rates were measured. Two values of $r$ are listed when CO and HCN measurements were not simultaneous.

$^\dagger$We assumed Q(HCN) $\sim$ Q(CN) for 29P, see \cite{wom17}.
\end{flushleft}
\medskip 

We briefly compare the CO and HCN production rates with other comets, since emission from these two species are commonly detected and their ratio may provide insights to the chemical composition of the nucleus and/or coma (Table 2). The comets identified by name are those reported to be CO-rich. Also listed in the table are average values for larger groups of comets, such as Oort Cloud, Jupiter Family, and finally an ``all comets" average value. As the table shows, the relative production rate value derived for R2 is extraordinarily high: Q(CO)/Q(HCN) $>$ 5000 at $r \sim$ 3 au. The average value for all comets measured is Q(CO)/Q(HCN) $\sim$ 25 and this ratio varies by less than a factor of three between Jupiter Family Comets (JFCs) and Oort Cloud Comets (OCCs). The only group where it noticeably departs from the average value is for CO-rich comets, which are listed individually in the top panel of the table. It is perhaps not surprising that the comets designated as CO-rich also have elevated Q(CO)/Q(HCN) values, but even among these CO-rich comets the limit derived for R2 is the highest values to date for any comet.


The very high Q(CO) and very low Q(HCN) in R2 is difficult to understand in terms of typical comet compositions. At 3 au, R2's comet nucleus has not received much solar heating and so it will preferentially release CO over HCN, due to its higher volatility. This behavior was seen in the abundance ratio of CO/HCN in Hale-Bopp, which decreased as the comet got closer to the Sun and more HCN was released (see \citet{biv02} and Table 2). There are not many measurements of both CO and HCN in comets at $\sim$ 3 au, but R2's value is substantially higher than those measured for the CO-rich comets Hale-Bopp, Christensen or C/2010 G2 Hill in the range of 2.5 - 3.0 au, suggesting that R2's high CO/HCN ratio cannot be explained solely due to volatility differences between the two molecules. Interestingly, the highest CO/HCN values were obtained in comets known to be both distantly active and CO-rich (R2, 29P, Christensen and Hale-Bopp). This is worth looking into further, but the data are sparse. Even for a comet at 3 au, the HCN upper limit that we derived is extraordinarily low. Another possible clue is that R2's coma is largely gaseous with very little dust \citep{coc18}, and this may be related to the significantly decreased amounts of HCN and other volatiles. The chemical composition of R2's coma is noticeably atypical when compared to other comets.

\medskip 
\subsection{High N$_2$ Production Rates}

Searching for additional clues to the unusual chemical composition of this comet, we now turn our attention to molecular nitrogen. Measuring cometary N$_2$ is of considerable importance for many reasons, including testing models of the condensation and incorporation of ices in the protosolar nebula, and calculating the N$_2$/NH$_3$ abundance ratio, which is a key diagnostic of primordial physical and chemical conditions \citep{feg89,wom92}. N$_2$ is also a highly volatile molecule and it can contribute to comet activity if significantly incorporated in the nucleus. Furthermore, N$_2$ is trapped and released in a manner similar to Argon, and thus, detecting N$_2^+$ emission in any coma suggests that Ar may also be present in high amounts \citep{owe95}. Despite its importance, it has been difficult to measure the N$_2$ abundances for all but a few comets \citep{lutz93,coch02,kor14,rub15,iva16}. Strikingly, N$_2^+$ optical emission is clearly detected in R2, with a measured abundance ratio of N(N$_2$)/N(CO)=0.06 \citep{coc18,coc18b}. 

The N$_2$/CO ratio is an important observational constraint for testing the formation environment for cometary ice; it is determined by the temperature of the gases when they were incorporated into the ice as well as any subsequent processing that may have preferentially affected one volatile over the other. In order to place this in context, we briefly review two general scenarios for comet formation: one where comets agglomerated from pristine amorphous water ice grains originating from the interstellar medium onto which N$_2$ and CO condensed in the protosolar nebula \citep{owen93}. This model proposes that the N$_2$/CO ratio in ices strongly depends on the temperature of the materials at the time the volatiles condensed or were trapped. The ratio derived from R2's optical spectra of N$_2^+$ and CO$^+$ agrees very well with the predicted value of N$_2$/CO = 0.06 for icy planetesimals forming in the solar nebula at about 50 K \citep{owe95,iro03}.  Alternately, comets may have agglomerated from crystalline water-ice grain clathrates that trap N$_2$ and CO. Due to its relatively small size, N$_2$ is not readily trapped by clathrates, which leads to a lower predicted ratio of N$_2$/CO ranging from $\sim$ 0.002 to 0.02 \citep{mou12}. Thus, perhaps the measured value of N$_2$/CO = 0.06 in R2 is not consistent with a clathrate model.  

In addition to being relatively rich in N$_2$, we note that this comet is severely depleted in HCN, as discussed in the previous section. Physicochemical models of nitrogen chemistry in protostellar disks show that photodissociation of N$_2$ leads to production of HCN \citep{hily17}. It is interesting to consider that comet R2 may have formed in the protosolar nebula disk where there was significant N$_2$ shielding that led to the high N$_2$ and decreased HCN abundances.

N$_2$ cannot undergo rotational transitions due to lack of a permanent dipole moment, and thus emits no radiation at millimeter-wavelengths. However, because the N$_2$ abundance can put such strong constraints on comet formation models, we derived an N$_2$ column density and production rate using the N$_2$/CO abundance ratio calculated from optical spectra and our CO results. We chose the CO data from 2017 Dec 22, because it is closest in time to the \cite{coc18b} N$_2$/CO value on 2017 Dec 8-10. We derived an N$_2$ column density of N(N$_2$) = (1.1 $\pm$ 0.2)x10$^{13}$ cm$^{-2}$ and a production rate of Q(N$_2$) = (2.8 $\pm$ 0.4)x10$^{27}$ molecules s$^{-1}$. This production rate corresponds to a mass loss rate of  130 kg s$^{-1}$  for N$_2$. Determining the N$_2$ production rate in this manner can be useful in comparison with those of water and/or NH$_3$, if these volatiles' abundances are established through direct observation or via their daughter products \citep{teg92,cro89}. 


\subsection{Conclusions\label{subsec:conclusion}}

We report the first detection of neutral CO emission, and an upper limit of HCN, in comet C/2016 R2 Pan-STARRS at $r \sim$ 3 au. The CO line profile shape is characteristic of CO emission seen in other comets at this distance, with a linewidth of $\sim$ 0.8 km s$^{-1}$ and is slightly blue-shifted from the ephemeris velocity. A 64\arcsec x 64\arcsec~ map ($\sim$ 96,000 km x 96,000 km at the comet's projected distance) shows that the CO emission peaks in intensity at the ephemeris position and decreases by 20-40\% at the off-centered positions. The spectra and map are consistent with CO arising from a combination of an isotropic source and an active area on the sunward side.

If comet R2's CO output is proportional to surface area of the nucleus, then we find that the radius need not exceed $R_{R2}$ $\sim$ 15 km in order to explain the measured CO production rate. Thus, R2 may be larger-than-average in size, but need not be a giant comet in order to explain the measured CO production rates. 

The very large amount of CO, and the apparent absence, or very low outgassing rate, of HCN, leads to a CO/HCN production rate ratio over 5000. This is remarkably high compared to other comets at $r$ = 3 au, even when including other comets known to be CO-rich. The high CO/HCN ratio cannot be explained solely due to volatility differences between the two, and may represent a compositional difference between R2 and most other comets. When considered along with the high outgassing rate of N$_2$, it is interesting to consider whether comet R2 formed in a region of the protosolar nebula with substantial N$_2$ shielding, which could have led to higher N$_2$ and decreased HCN abundances.

N$_2$ production rates were derived from the N$_2$/CO ratio \citep{coc18b} and our CO production rates, and were calculated to be Q(N$_2$) = (2.8 $\pm$ 0.4)x10$^{27}$ molecules s$^{-1}$ at 3 au. N$_2$ production rates will be valuable for comparison with those of water and/or NH$_3$, if detected in this comet. 

R2's coma composition is clearly very different from other comets observed thus far, both in the high N$_2$ abundance, and significant decrease in other typically abundant molecules, such as HCN. Further observations of this comet along all spectral ranges are highly encouraged, especially those of NH$_2$ in the optical, and NH$_3$, CO$_2$ and CH$_4$ in the infrared in order to measure the key diagnostic ratios N$_2$/NH$_3$, CO/CO$_2$ and CO/CH$_4$, which will provide observational tests for formation models of the environment of this comet. 
\medskip 

\acknowledgments

M.W. and K.W. acknowledge support from NSF grant AST-1615917. We also thank the ARO 10m SMT crew. The SMT is operated by the ARO, the Steward Observatory, and the University of Arizona, with support through the NSF University Radio Observatories program grant AST-1140030. This work was completed with the GLIDASS CLASS software: http://www.iram.fr/IRAMFR/GILDAS

%

\vspace{5mm}
\facilities{Arizona Radio Observatory (SMT)}

\newpage



\end{document}